\documentclass[osajnl,showpacs]{revtex4}
\usepackage[applemac]{inputenc}
\usepackage{amsfonts}
\usepackage{amsmath}
\usepackage{graphicx}
\usepackage{color}
\usepackage{bm}
\definecolor{pink}{rgb}{1,1,0} 
\definecolor{yellow}{rgb}{1,1,0}
\definecolor{orange}{rgb}{1,0.5,0}
\definecolor{white}{rgb}{1,1,1}

\def\ifcorrige{\iffalse}

\begin{document}

\title{Coexisting ordinary elasticity and superfluidity in a 
model of defect-free supersolid}

\author{
Christophe Josserand$^1$, Yves Pomeau$^2$ and Sergio Rica$^{2,3}$}
\affiliation{$^1$Laboratoire de Mod\'elisation en M\'ecanique,
CNRS UMR 7607, Case 162, 4 place Jussieu, 75005 Paris, France\\
$^2$Laboratoire de Physique Statistique, ENS-CNRS, 24 Rue Lhomond, 75005 Paris, France.\\
$^3$ Departamento de F\'\i sica, Universidad de Chile, Blanco Encalada 2008, Santiago, Chile.}

\begin{abstract}
We present the mechanics of a model of supersolid in the frame of the Gross-Pitaevskii equation at $T=0K$ that do not require defects nor vacancies. A set of coupled nonlinear partial differential equations plus boundary conditions is derived. The mechanical equilibrium is studied under external constrains as steady rotation or external stress. Our model displays a paradoxical behavior: the existence of a non classical rotational inertia fraction in the limit of small rotation speed and no superflow under small (but finite) stress nor external force. The only matter flow for finite stress is due to plasticity. 
\end{abstract}

\maketitle

The recent surge of interest in supersolids
\cite{chan} makes it important to reach a clearer 
understanding 
of the mechanical properties of such materials. In particular why a supersolid behaviour is observed in a rotating experiment, whilst, as in a ordinary solid no constant mass flux is driven by pressure gradient? (see Refs. \cite{pressure})   In \cite{pomric} two of 
us (YP and SR) proposed a 
fully explicit 
model of supersolid where many properties can be discussed in details. We thought timely to reconsider 
this model, in particular with respect to its properties of elasticity 
coupled to its ability to carry some sort of superflow in the absence 
of any 
defect. Although supersolidity is often related to the presence of defects, vacancies and so forth our model introduces an important 
difference between ordinary (classical) crystals and supersolids: 
in perfect classical crystals there is an integer  number (or a simple 
fraction) of atoms per unit cell. Therefore the number density 
and the lattice parameters are not independent. On the 
contrary, in our model of supersolid, there is no such 
relation. The lattice parameters and the average density can be 
changed independently. 

 This model 
is based upon the original Gross-Pitaevksii equation (G--P later)  \cite{pitgross}  with an integral term with a kernel that can be seen as a two-body potential in a first Born approximation\footnote{This is an exact approximation for the full quantum problem in the van der Waals limit of long range low amplitude potential $U(r) = \gamma^3 {\mathcal U} (\gamma r)$ whenever $\gamma\rightarrow 0$. }.
This model yields the exact spectrum found long ago by Bogoliubov  \cite{bog}, 
namely a relation between the energy of the elementary excitations and 
their momentum depending on the two-body potential. In this framework the roton minimum is a 
precursor of crystallisation. Something predicted in \cite{enz} where the possibility of a linear 
instability was only considered, although the transition is subcritical-first 
order \cite{pomric}. By increasing the 
density, crystallization happens through a first order phase transition (see figure \ref{grav}). As
shown in \cite{pomric} the crystal phase shows a periodic modulation 
of density in space together with some superfluid-like behaviour under rotation. 

The aim of the present letter is to 
show that, besides this behaviour, the system has also solid-like 
behaviour, at least under small stress. At larger stress, it 
flows plastically, the plasticity being facilitated by the eventual presence 
of defects. 
We derive the 
equation of motion for  the average density $n$, the phase $\Phi$ 
and the displacement $\bm{u}$ in the solid. A new propagating mode 
appears besides the usual 
longitudinal and transverse phonons in regular crystals. This mode 
is partly a modulation of the coherent quantum phase, like the phonons in superfluids at zero temperature. 
We discuss at the end the boundary conditions and how to handle steady rotation and pressure driven flow in this 
 model. 
 
Our starting point is the original G--P
 equation \cite{pitgross} for the complex valued wavefunction $\psi({\bm{r}},t)$ common to all bosonic particles of mass $m$: 
\begin{equation}
    i\hbar \frac{\partial \psi}{\partial t}= 
    -\frac{\hbar^{2}}{2m}{\bm{\nabla}}^{2}\psi+ \psi\int{\mathrm{d}}{\bm{r}}' U({\bm{r}}' - 
{\bm{r}}) |\psi({\bm{r}'},t)|^2
    \mathrm{,}
\label{GP}
\end{equation}
where $U(\cdot)$ is a two body 
 potential depending on the distance. For the numerics we choose a potential $U(|\bm{r}|) = U_0 \theta(a-|\bm{r}|)$, with  $\theta(.)$ Heaviside function. 

The ground state is a solution of 
equation (\ref{GP}) of the form 
$\psi_{0}({\bm{r}})e^{-i\frac{E_{0} t}{\hbar}}$. It is a crystal when $\psi_{0}({\bm{r}})$ is a periodic function such that 
$\psi_{0}({\bm{r}}+ q_{a} {\bm{a}}+ q_{b} {\bm{b}}+q_{c}{\bm{c}}) = \psi_{0}({\bm{r}})$ 
for $q_{a,b,c}$ arbitrary integers, ${\bm{a}}\mathrm{,}{\bm{b}}\mathrm{,}{\bm{c}}$ being vectors defining the elementary lattice cell. This 
solution is the ground state in the sense that, given an 
average number density $ n = \frac{1}{\Omega}\int \mathrm{d}{\bm{r}} 
|\psi_{0}|^2$, $\Omega$ being the total volume, the lattice parameters and the 
function $\psi_{0}({\bm{r}})$ make $E_{0}$ the 
smallest possible. 
The local density 
$n({\bm{r}}, t)$, the displacement field ${\bm{u}}({\bm{r}},t)$ of the 
crystal lattice and the slowly varying 
phase $\Phi({\bm{r}},t)$ of $\psi(t, \bm{r})$. 

The Lagrangian density for the G--P equation (\ref{GP}) reads in polar coordinates, $\psi= \sqrt{\rho} e^{i\phi}$: 
\begin{equation}
    {\mathcal{L}} =-\hbar \rho \frac{\partial 
       \phi}{\partial t}- 
       \frac{\hbar^{2}}{2m}\left({\rho} ({\bm{\nabla}}\phi)^2 + 
       \frac{1}{4\rho}({\bm{\nabla}}\rho)^2 \right) - \\ \nonumber
       \frac{1}{2} \rho({\bm{r}})  \int 
       {\mathrm{d}}{\bm{r}}' U({\bm{r}}'-{\bm{r}}) \rho({\bm{r}}') 
       \mathrm{.}
   \label{eq:Lfull}
    \end{equation}
The ground state is given by the solution of the nonlinear 
    integro-differential equation 
    for $\rho$ derived by variation of the action  taking the phase field 
    $\phi$
    uniform in space: $\phi = -\mu t/\hbar$, $\mu$ constant, that is a solution of: 
    \begin{equation}
	   -\mu + \frac{\hbar^{2}}{4m}\left(
	   \frac{({\bm{\nabla}}\rho)^2}{2\rho^2} - \frac{\nabla^2 \rho}{\rho} \right) + \int 
	   {\mathrm{d}}{\bm{r}}' U({\bm{r}}'-{\bm{r}}) \rho({\bm{r}}') = 0
	   \mathrm{.}
       \label{eq:LstatEulL}
	\end{equation}
This ground state solution, if periodic in space, as we shall assume in full agreement with our numerical results, depends on 
the dimensionless parameter $\Lambda=  U_0\frac{m a^2}{\hbar^2} n a^3$ only \cite{pomric}. 
Although in Ref. \cite{pomric} we discussed a 
ground state as a modulation close to an uniform  density near the transition, that is for a finite roton gap, we have observed numerically that a crystal ground  state exists in a wide range of densities. In the limit $\Lambda \gg 1$ the lattice tends to an array of 
sharp density pulses distant of $a$, the width of the pulse decreasing like $\Lambda ^{-1/2}$. 

Let $\rho_{0}({\bm{r}}|n)$ be a ground state solution, then  $\rho_{0}({\bm{r}}-{\bm{u}}|n)$ is also a ground state solution with the same $\mu$, for a constant displacement field ${\bm{u}}$. 
The general perturbations around the ground state allow that $\Phi$, $\bm u$ and $n$ become slow varying fields on space and time. As in Ref. \cite{pomric} we follow 
the general method called homogenization \cite{homogenization}. This splits 
cleanly the long-wave behaviour of the various parameters and the 
short range periodic dependence upon the lattice parameters. Let write the {\it Ansatz} for density and phase: 
\begin{eqnarray}
&\rho = \rho_0({\bm r}-{\bm u} ,n({\bm r},t) ) + \tilde\rho({\bm r}-{\bm u} ,n ,t)+\dots\nonumber\\
&\phi= \Phi({\bm r},t)  + \tilde\phi({\bm r}-{\bm u} ,n ,t)+\dots \label{ansatz}
\end{eqnarray} where $\Phi$, $\bm u$ and $n$ are slow varying fields and $\tilde\phi$ and $ \tilde\rho$ are small and fast varying periodic functions. Introducing this {\it Ansatz} into the Lagrangian (\ref{eq:Lfull}) one gets an effective Lagrangian made of four kind of terms:

{\it i)} As $n$
changes continuously the periodic solution of the integrodifferential equation 
(\ref{eq:LstatEulL}), say $\rho_{0}(\bm{r})$,  can be considered as a regular function 
of the Lagrange multiplier $\mu$ imposing the average density $n$. Therefore 
by integrating over an unit cell ($V$) of the lattice the Lagrangian from which (\ref{eq:LstatEulL}) 
is derived one obtains an averaged Lagrangian that depends on $n$ only that writes
\ifcorrige
\begin{multline}
  - {\mathcal L}_n=    {\mathcal E}(n) =   \frac{1}{V}\int_V{\mathrm{d}}{\bm{r}}
	  (\frac{\hbar^{2}}{8m \rho_{0}}({\bm{\nabla}}\rho_{0})^2  -\mu \rho_{0} + \\
	  \frac{1}{2}\rho_{0}({\bm{r}}) \int 
	  {\mathrm{d}}{\bm{r}}' U({\bm{r}}'-{\bm{r}}) \rho_{0}({\bm{r}}') )
	  \mathrm{.}
      \label{eq:Lstatn}
       \end{multline}
       \fi
\begin{equation}
  - {\mathcal L}_n=    {\mathcal E}(n) =   \frac{1}{2V}\int_V {\mathrm{d}}{\bm{r}} \rho_{0}({\bm{r}}) \int U({\bm{r}}'-{\bm{r}})  \rho_{0}({\bm{r}'}) {\mathrm{d}}{\bm{r}'}
	  \mathrm{.}
      \label{eq:Lstatn}
       \end{equation}
This yields the simplest case of homogenization.

 {\it ii)} Similarly, terms mixing the slow varying phase field $ \Phi({\bm r},t) $ and $\rho_{0}(\bm{r})$ can be averaged directly leading to
      $    {\mathcal{L}}_{\Phi} =-n\left(\hbar \partial_t\Phi  +  \frac{\hbar^2}{2m}  ({ \bm \nabla} \Phi)^2 \right) $ where 
$ n = \frac{1}{V} \int_{V}  \rho_{0}({\bm r})\, d{\bm r}. $

Next contributions need to solve the Euler-Lagrange conditions for the fast variables  $\tilde\phi$ and  $\tilde\rho$. 
We shall sketch the effective Lagrangian for the phase $\tilde\phi$, leaving for the reader a similar calculation for the deformation part.

{\it iii)} The $\tilde\phi$ dependence term of this Lagrangian can be re-written of the form:
$
 {\mathcal L}_{\tilde\phi} =    - \frac{\hbar^2}{2m}  \int \left( 2  \rho_{0}  {\bm{A}}\cdot{\bm{\nabla}}\tilde\phi+ 
   \rho_{0} ({\bm{\nabla}}{\tilde{\phi}})^2\right)  \, 
      d{\bm r}  \mathrm{,}
$  
where $ {\bm A } = \left({ \bm \nabla} \Phi- ({ \bm \nabla} \Phi\cdot { \bm \nabla})  {\bm u}  -   \frac{m}{\hbar}\partial_t{\bm u}\right)$ (considered as a constant in the unit cell). The  Euler-Lagrange condition for $  {\mathcal L}_{\tilde\phi} $ reads
$
    {\bm{A}}\cdot{\bm{\nabla}}\rho_{0} + 
    {\bm{\nabla}}\cdot\left(\rho_{0}{\bm{\nabla}}{\tilde{\phi}}\right) = 
    0
    \mathrm{.}
 $
This Poisson-like equation is to be solved within the unit cell of the  lattice, for a function $\tilde\phi$ that is periodic with  the same period as $\rho_{0}$. The result (that can be expressed as the minimum of a certain Rayleigh-Ritz functional)
 is linear in $\bm{A}$ and can be written as $\tilde\phi = {{K}}_{i} A_{i}$ where ${\bm{K}}({\bm r})$ is a vector-valued function of ${\bm{r}}$ that is periodic and satisfies $\nabla_i\rho_{0} + 
    {\bm{\nabla}}\cdot\left(\rho_{0}{\bm{\nabla}} K_i\right) = 
    0.$ Putting the result into the Lagrangian $
 {\mathcal L}_{\tilde\phi}$ one obtains the relevant contribution for the slowly varying part of the phase: 
$
 {\mathcal{L}}_{\tilde\phi} = 
  \frac{\hbar^2}{2m} {\varrho}_{ij} A_iA_j  $
with  the positive defined matrix 
$   {\varrho}_{ij} = \frac{1}{V} \int_{V}  \rho_{0}({\bm r})\, 
   {\bm \nabla} {K}_{i}\cdot  
    {\bm \nabla} K_{j}   \, d{\bm r}  \, .
$ The Lagrange function $ {\mathcal{L}}_{\tilde\phi}$ depends on the slow variables only. We shall restrict ourselves below to crystal structures sufficiently symmetric to make $\varrho_{ij}$ diagonal $\varrho_{ij} = \varrho(n) \delta_{ij} $.
The quantity $\varrho(n)$ is zero if the crystal modulation is absent and would be very small for Bose-Einstein condensate with a non local interaction term. $\varrho(n)\rightarrow n$ when all the mass is strongly localized  in the center of the cell site with a small overlap in between the different sites. This is presumably the situation of almost all materials in their solid state at low temperature. A large Young modulus is likely a measure of the small overlap of the wave functions from one site to the next, making $^4$He exceptional at this respect.  In other words when $\varrho(n)\rightarrow n$ the supersolid behaves as a ordinary solid state.

{\it iv)} The same method of homogenization works for the long-wave 
    perturbations of gradients of the displacement ${\bm{u}}$ and yields a contribution to the Lagrangian that reads 
   $ {\mathcal{L}}_{\bm u} = -\frac{1}{2} \lambda_{ijkl}  \frac{\partial u_{i}}{\partial x_{j}}\frac{\partial 
    u_{k}}{\partial x_{l}} 
   \mathrm{.}$
The coefficients $\lambda_{ijkl}$ are given by integrals over the unit cell of various functions defined explicitely. This is the familiar elastic energy of a Hookean solid.
   
To summarize, the effective Lagrangian reads: 
\begin{equation}
 {\mathcal L}_{eff} =- \hbar n \frac{\partial \Phi}{\partial t}-	  \frac{\hbar^2}{2m}\left[n\left({\bm \nabla}\Phi\right)^2 -\varrho(n)
	  \left({\bm \nabla}\Phi - \frac{m}{\hbar}  \frac{{\mathrm D} {\bm u}}{{\mathrm D}t} \right)^2\right]      -{\mathcal E}(n) - \frac{1}{2}\lambda_{ijkl}  \frac{\partial u_{i}}{\partial x_{j}}\frac{\partial u_{k}}{\partial x_{l}}    \label{eq:lagrangtimedeptotal}
      \end{equation}
where
    $
	  \frac{{\mathrm{D}}{\bm{u}}}{\mathrm{D}t} = 
	  \frac{\partial{\bm{u}}}{\partial t} + 
	  \frac{\hbar}{m}{\bm{\nabla}}\Phi\cdot \bm {\nabla} \bm{u}
	  \mathrm{.}
	  $
This expression is remarkable because it is fully explicit for a given ground 
state of the G–P model. 
As one can check this Lagrangian is Galilean invariant.

We conjecture that, because this Lagrangian satisfies the 
symmetries imposed by the underlying physics and because it includes 
a priori all terms with the right order of magnitude with respect to the 
derivatives, the general Lagrangian of any supersolid at zero 
temperature has the same structure. In a recent paper, Son \cite{son} derives a Galilean invariant Lagrangian such that (\ref{eq:lagrangtimedeptotal}) is a sub-class but with well defined coefficients like $\varrho(n)$, 
${\cal E} (n)$ and $\lambda_{ijkl}$ depending on the details of the crystal 
structure.

The dynamical equations are derived by variation of the action $\int {\mathcal L} d^3{\bm r}\, dt $
which is seen as a functional of $n$, $\Phi$ and $\bm{u}$. \ifcorrige The final result is a set of coupled partial differential 
equations for the those fields.\fi The variation with respect to $n$, $\bm{u}$ and $\Phi$ yields (writing $\varrho'(n) = d\varrho/dn$, {\it etc.}): 

\begin{eqnarray}
\hbar \frac{\partial \Phi}{\partial t} + 
\frac{\hbar^2}{2m}\left[\left({\bm{\nabla}}\Phi\right)^2 
-\varrho'(n)
\left({\bm \nabla}\Phi - \frac{m}{\hbar}  \frac{{\mathrm D} {\bm u}}{{\mathrm D}t}\right)^2
\right] + 
  {\cal E} '(n) + 
	   \frac{1}{2}
	  \lambda'_{ijkl}\frac{\partial u_{i}}{\partial 
		x_{j}}\frac{\partial 
		u_{k}}{\partial x_{l}} & = & 0
		  \mathrm{.} 
\label{eq:lagrangtimedeptotalBernoulli} \\
 m \frac{\partial}{\partial t}\left[
   \varrho(n) (   \frac{{\mathrm D}u_i }{{\mathrm D}t}  - \frac{\hbar}{m} \frac{\partial  \Phi}{\partial 
      x_{i}} )\right]  -  \frac{\partial}{\partial x_{j}}\left(\lambda_{ijkl}\frac{\partial u_{k}}{\partial 
		x_{l}}\right) 
    + \hbar \frac{\partial}{\partial x_{k}}\left[\varrho ( \frac{{\mathrm D}u_i}{{\mathrm D}t} -
      \frac{\hbar}{m}\frac{\partial 
      \Phi}{\partial x_{i}}  )\frac{\partial 
      \Phi}{\partial x_{k}} \right]  &= &0
		  \mathrm{.}
\label{eq:lagrangtimedeptotalcauchy}\\
 \frac{\partial n}{\partial t} +  \frac{\hbar}{m}{\bm{\nabla}}\cdot\left( n  {\bm{\nabla}}{\Phi} \right) -
\frac{\hbar}{m}\partial_k \left( \varrho(n) (\delta_{ik} - \partial_k u_i)\left(\partial_i {\Phi} -
	\frac{m}{\hbar}\frac{\mathrm{D}u_i}{\mathrm{D}t}\right)\right) &=& 0
				 \mathrm{.}
   \label{eq:lagrangtimedeptotalmass}
\end{eqnarray}	
The latter equation reduces to the familiar equation of mass conservation 
for potential flows whenever $\varrho(n) = 0$, namely in the absence of 
modulation of the ground state. Although our equations of motion  (\ref{eq:lagrangtimedeptotalBernoulli},\ref{eq:lagrangtimedeptotalcauchy},\ref{eq:lagrangtimedeptotalmass}) and the one of Andreev--Lifshitz  are almost identical in the zero temperature limit (see eqns. (16) of Ref.  \cite{andreev}), our model
has significant differences with their. Our solid cannot be considered as the normal component
of a two ``fluids'' system, because it is on the same footing (phase coherent) as the superfluid  part at
$T=0K$. Therefore, at small finite temperature, our model has a normal component that is a
fluid of vanishing density at $T=0K$, besides its coherent superfluid and solid part and should change the superfluid density fraction. Following Landau's ideas, this normal fluid is a gas of 
quasi-particles with the mixed spectrum able to carry momentum whilst
the coherent part (superfluid plus solid) stays at rest.

The Euler-Lagrange conditions impose also the boundary conditions for the equations of 
motion:
$$
	\frac{\hbar}{m} \left( n  {\partial_k}{\Phi}  -
	\varrho  (\delta_{ik} - \partial_k u_i)\left(\partial_i {\Phi} -
	\frac{m}{\hbar}\frac{\mathrm{D}u_i}{\mathrm{D}t}\right) \right) \hat e_k =  n V_k \hat e_k \mathrm{.}
$$
where $V_k$ is the local speed of the solid wall of the container and $ \hat e_k$ is normal to it. The displacement moves with the wall: $\frac{\mathrm{D}{\bm u}}{\mathrm{D}t}  = \bm V$.

Let us look at small perturbations around a nondeformed ($\bm u=0$) and steady ($\bm \nabla \Phi =0$) state of average density $n$.
The linearized version of (\ref{eq:lagrangtimedeptotalBernoulli},\ref{eq:lagrangtimedeptotalcauchy},\ref{eq:lagrangtimedeptotalmass}) shows that the shear waves are decoupled from the compression and phase (Bogoliubov-like) waves. The dispersion relation for the coupled compression and phase waves leads to a simple algebraic equation. 
 In the limit $\varrho(n)\rightarrow 0$ the crystal structure disappears and the phase mode propagates at the usual speed of sound found by Bogoliubov $c = \sqrt{\mathcal E''(n)/(m n)}$.
In the limit $\varrho(n)\rightarrow n$, that is whenever the supersolid behaves as a regular solid state, the two propagation speeds are ($c_K$ is the longitudinal elastic wave speed \cite{landau})
$v_1  = \sqrt{ c_K^2+c^2}  $ and $  v_2 = \sqrt{c_K^2c^2 /(c_K^2+c^2) }   \sqrt{1-\varrho(n)/n} $ meaning that the phase mode disappears at the transition supersolid-solid. 

As suggested by Leggett \cite{leggett} an Andronikashvili kind of experiment could manifest a non classical rotational inertia (NCRI). Indeed let us supose that the wall of the container of volume $\Omega$ rotate with an uniform angular speed $\omega$. Then for low angular speed the crystal moves rigidly with the container $\dot {\bm u} = {\bm \omega}\times {\bm r}$ without any elastic deformation. The densities $n$ and $\varrho(n)$ being constant in space, equation (\ref{eq:lagrangtimedeptotalmass}) simplifies into 
\begin{equation}\nabla^2\Phi=0\, {\rm in} \, \Omega \,{\rm with} \, {\bm \nabla} \Phi \cdot\hat e = (m/\hbar) ({\bm \omega}\times {\bm r}) \cdot \hat e \, {\rm on} \, \partial\Omega .\label{perfect} \end{equation}
This mathematical problem (\ref{perfect}) has an unique solution \cite{milne}. The moment of inertia comes directly from the energy per unit volume of the system: $E = \Phi_t \frac{\delta {\mathcal L}}{\delta \Phi_t } + {\bm u}_t \cdot \frac{\delta {\mathcal L}}{\delta {\bm u} _t } -{\mathcal L}.$ In the rotating case $E = \frac{1}{2} I_{ss} \omega^2$ where $I_{ss}$ is the $zz$ component of the inertia moment :
$
I_{ss} = m (n-\varrho(n) ) {\mathcal I}_{pf} + m\varrho(n) {\mathcal I}_{rb}$ 
with 
${\mathcal I}_{pf} = \int_\Omega (\bm \nabla \Phi)^2 d{\bm r}$, $\Phi$ solution of (\ref{perfect}), $\omega$, $m$ and $\hbar$ taken to 1. This number depends on the geometry only,  ${\mathcal I}_{rb} $ is also a geometrical factor corresponding  to rigid body rotational inertia ($x\& y$ orthogonal to the axis of rotation)  ${\mathcal I}_{rb}= \int_\Omega(x^2+y^2) d{\bm r}$. The relative change of the moment of inertia whenever the supersolid phase appears is (here ${I}_{rb}=m n {\mathcal I}_{rb}$)
\begin{equation}(I_{ss}-I_{rb})/I_{rb} = - (1-\varrho(n)/n) (1-{\mathcal I}_{pf}/{\mathcal I}_{rb})\label{ncri}\end{equation}
Because ${\mathcal I}_{pf}<{\mathcal I}_{rb}$, one has $ (I_{ss}-I_{rb})/I_{rb} \leq 0$ as expected and observed experimentally \cite{chan}. The NCRI fraction (NCRIF) disappears, as well as the phase mode sound speed, when the supersolid phase recovers the ordinary solid phase ($ \varrho(n) \rightarrow n$).

\begin{figure}[hc]
\begin{center}
\centerline{{\it a} \includegraphics[width=8cm]{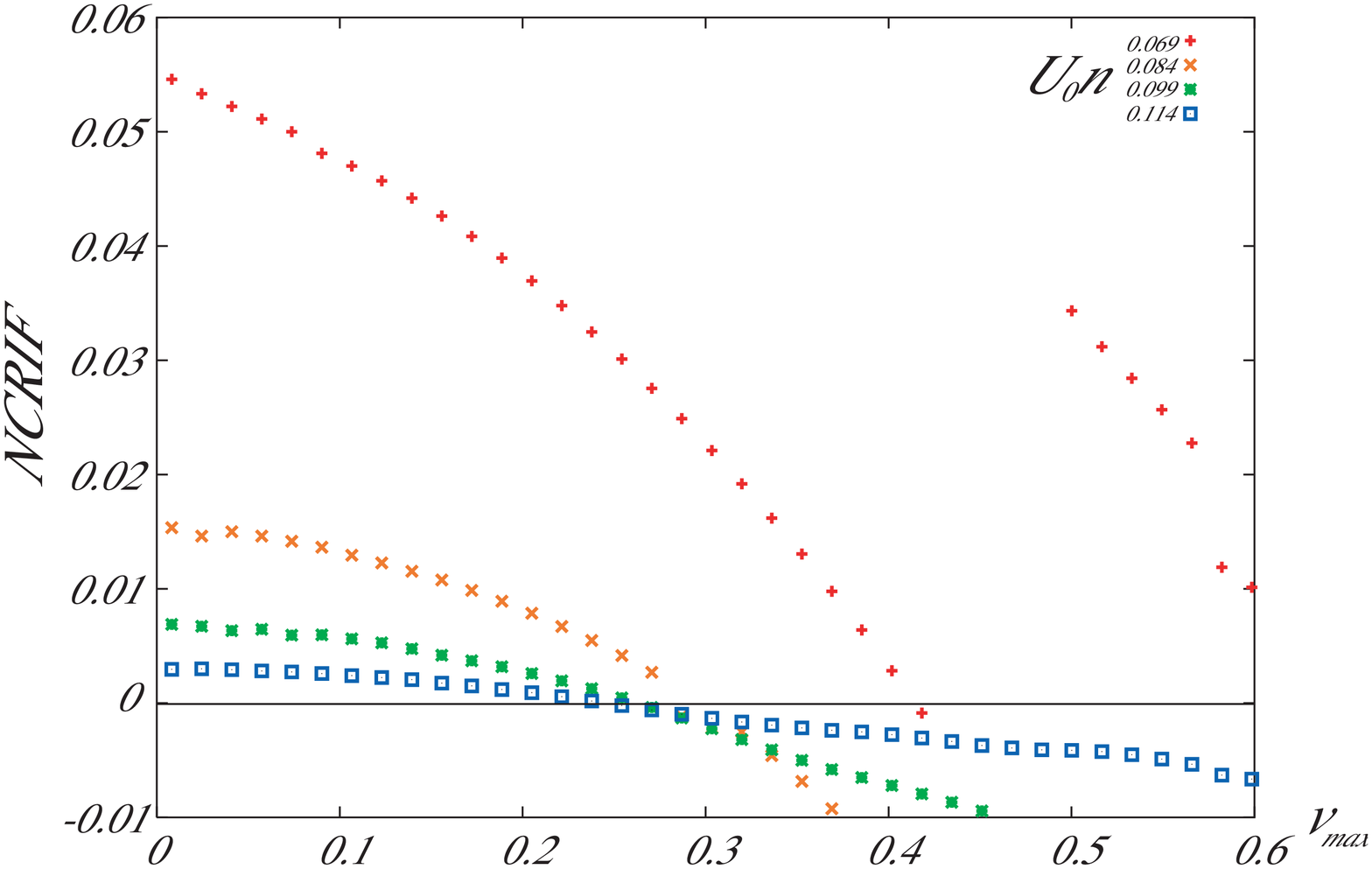} {\it b} \includegraphics[width=8cm]{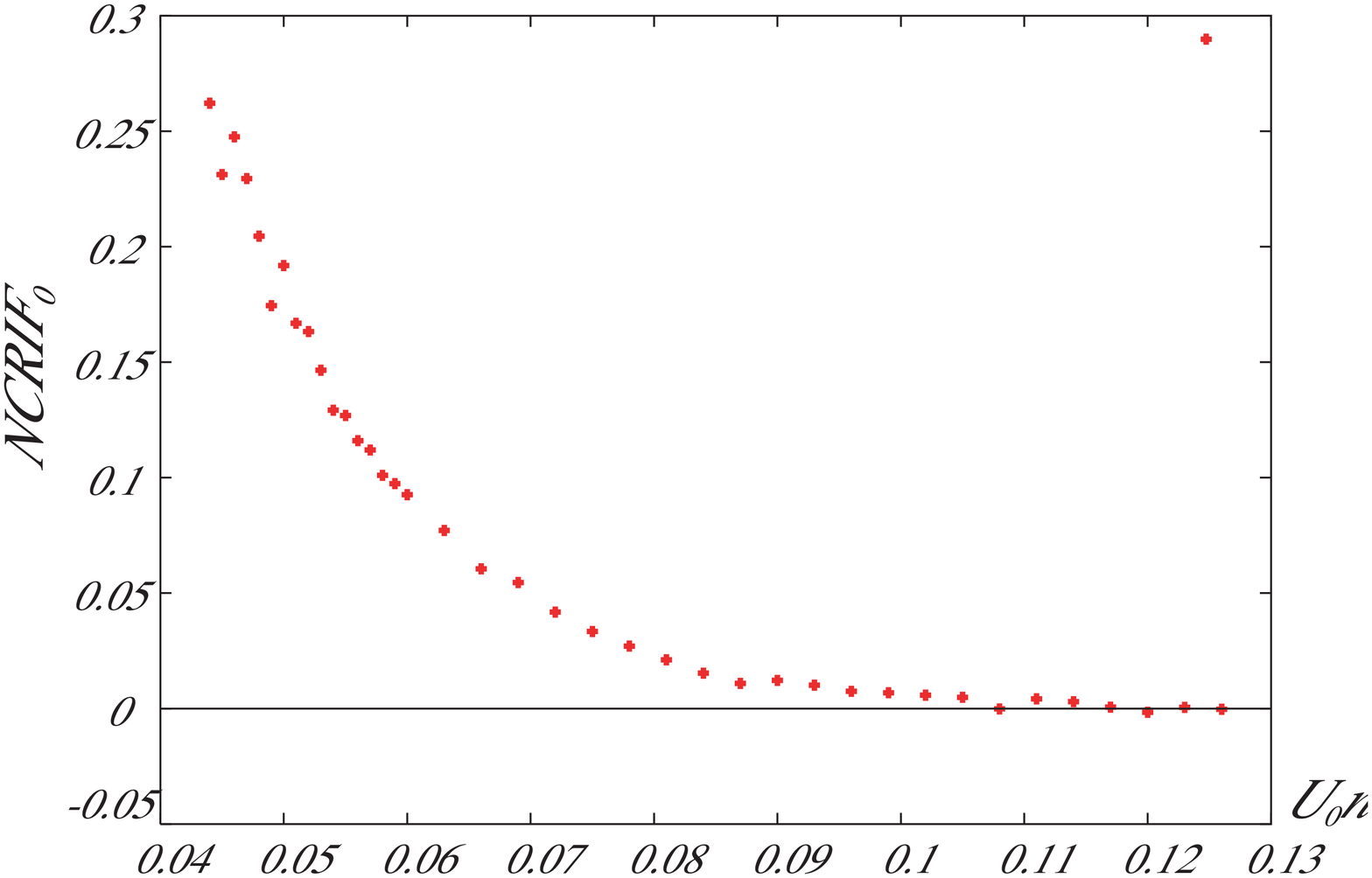}}
\caption{ \label{fig1} We implement a relaxation algorithm in Fourier space with $128\times 128$ modes to find a local minima  in a square cell of $96 \times 96$ units, the potential range $a = 4.3$, for different values of $U_0 n$. {\it a)} The $NCRIF \equiv 1-L_z'(\omega)/\left<I_{rb}\right>$ {\it vs.} the local Maximum speed $v_{max} = \omega L/\sqrt{2}$ for different values of the compression parameter $n U_0= 0.069, 0.084, 0.099 \&
0.114$ Here $\left<I_{rb}\right>$ is the average inertia moment for large $\omega$ computed numerically.
 {\it b)} The NCRIF$_0$ as a function of $nU_0$. Note that  {\it a)} and  {\it b)}  almost do not depend on the box size. }
\end{center}
\end{figure}

\begin{figure}[hc]
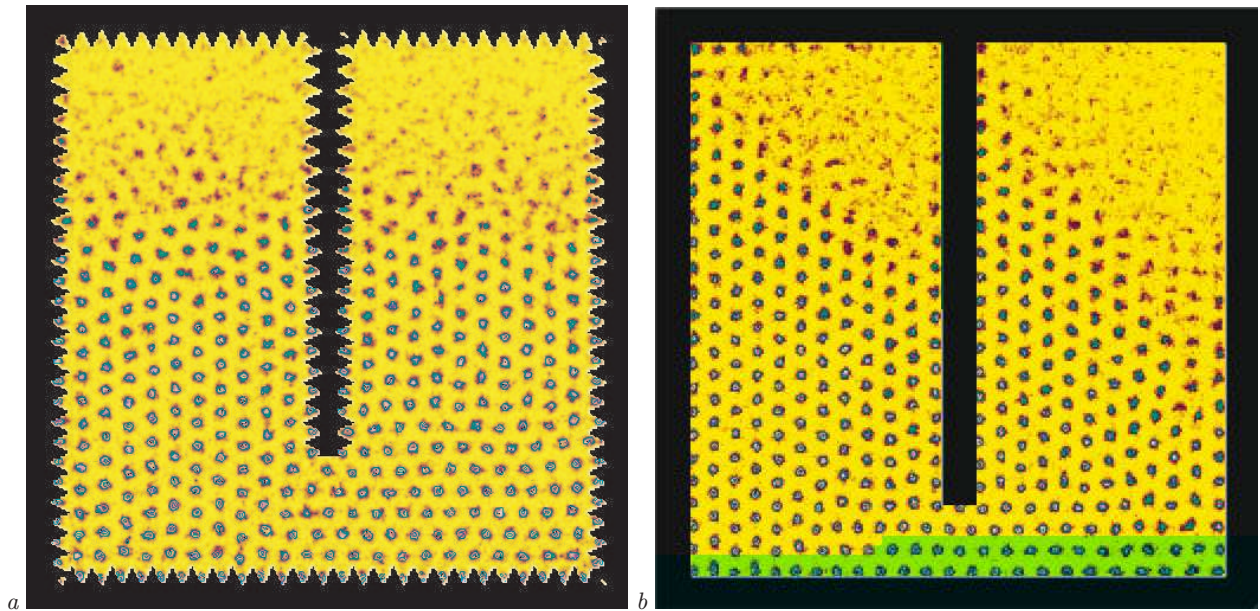

\begin{center}
\centerline{{\it a} \includegraphics[width=8cm]{fig2a.epsi} {\it b} \includegraphics[width=8cm]{fig2b.epsi}}
\caption{ \label{grav} We plot the density modulations $|\psi|^2$ (the dark points means a large mass concentration) of a numerical Simulation of eqn. (\ref{GP}) with Dirichlet boundary conditions with the shape of an u-tube as in the figure. We use a Crank-Nicholson scheme that conserves  the total energy and mass. The mesh size is $dx=1$, the nonlocal interaction parameters are chosen as $U_0 = 0.01$ and $a=8$ (physical constants $\hbar$ and $m$ are 1), finally the initial condition is an uniform solution $\psi= 1$ plus small fluctuations. One gets a crystalline state after $500$ units time;  then a vertical gravity of magnitude ${\mathcal G} =0.01$ is switched-on, and the system evolves for 500 time unites up to {\it a.} Then gravity orientation is tilted in 45$^\circ$. After 2000 time units the system evolves to a stationary situation {\it b} showing that the mass flow is only a transient. 
}
\end{center}
\end{figure}

Within the model presented here it is easy to implement a numerical procedure to put in evidence a NCRI in a 2D system. We shall first minimize $H-\omega L_z$  for different values of the angular frequency $\omega$, where $H=  \frac{\hbar^{2}}{2m}\int |\bm{\nabla}\psi|^2  {\mathrm{d}}{\bm{r}} +\frac{1}{2}\int U({\bm{r}}' - 
{\bm{r}})|\psi({\bm{r}},t)|^2 |\psi({\bm{r}'},t)|^2 {\mathrm{d}}{\bm{r}} {\mathrm{d}}{\bm{r}}' $ is the Hamiltonian and $L_z =  \frac{i\hbar}{2} \int  ( \psi^* {\bm r}\times {\bm\nabla}\psi -\psi {\bm r}\times {\bm\nabla}\psi^* ){\mathrm{d}}{\bm{r}}$
the angular momentum. The minimization should constrain a fixed total mass : $N =\int  | \psi|^2 {\mathrm{d}}{\bm{r}}$. Starting with $\omega=0$ one finds the minimizer and then by increasing $\omega$ step by step together with the minimization procedure we follow the evolution of the local minima. Figure \ref{fig1}-{\it a} represents the NCRIF as function of $\omega$, for different values of  $n U_0$. We observe a non-zero NCRIF in particular in the limit $\omega \rightarrow 0$. Fgure \ref{fig1}-{\it b} shows this limit NCRIF$_0$ as a function of the dimensionless compression $\Lambda=  U_0\frac{m a^2}{\hbar^2} n a^3$. Both curves are in qualitative agreement with recent experiments (see Fig. 3-D of \cite{chan}b and
Fig. 4 of \cite{chan}c).
Finally, we study a 
gravity (or pressure) driven supersolid flow. As early suggested by Andreev {\it et al.} \cite{andreev2} an experiment of an obstacle pulled by gravity in solid helium could be a proof of supersolidity. Different versions of this experiment failed to show any motion \cite{pressure}, therefore a natural question arises: How we can reconcile the NCRI experiment by Kim and Chan and the absence of pressure or gravity driven flows? 

In fact, our supersolid model (and it seems that supersolid helium too) reacts in different ways under a small external constrain such as stress, bulk force or rotation in order to satisfies the equation of motion and the boundary conditions. For instance, if gravity (or pressure gradient) is added then the pressure ${\mathcal E}'(n)$ balances the external  ``hydrostatic'' pressure $mg z$ in equation (\ref{eq:lagrangtimedeptotalBernoulli}) while the elastic behavior of the solid of equation (\ref{eq:lagrangtimedeptotalcauchy}) balances the external uniform force per unit volume $m n g$. No ${\bm \nabla} \Phi$ nor $\dot { \bm u}$ are needed to satisfy the mechanical equilibria. Moreover, a flow is possible only if the stresses are large enough to display a plastic flow as it happens in ordinary solids. 
In \cite{pomric} we showed that a flow around an obstacle is possible only if defects are created in the crystal, in this sense we did observe a plastic flow, however in the same model  we observe a ``superfluid-like''  behaviour under rotation without defects in the crystal structure. Indeed for a small angular rotation the elastic deformations come to order $\omega^2$ while ${\bm \nabla}\Phi$ or $ \dot { \bm u}$ are of order $\omega$, the equations of motion together with the boundary conditions leads to a NCRIF different from zero.

  We have realized a numerical simulation to test the possibility of a permanent gravity flow for different values of the dimensionless gravity ${\mathcal G}=\frac{m^2ga^3}{\hbar^2}$.
 Let us consider an U-tube as in Fig.\ref{grav}.
The system is prepared for $500$ time units into a good quality (but not perfect) crystalline state. A vertical gravity of magnitude $\mathcal G$ is switched-on and the system evolves for 500 time units more up to a new equilibrium state (see Fig. \ref{grav}-{\it a}).  The gravity is then tilted (with the same magnitude) at a given angle. 
A mass flow is observed at the begining from one reservoir into the other, but both vessels do not reach the same level eventually (see Fig. \ref{grav}-{\it b}). 
There is some dependence of the transferred mass on $\mathcal G$
till $\mathcal G \approx 0.0005$  and the mass transfer becomes negligible from fluctuations for $\mathcal G < 0.00025 $ indicating the existence of a yield-stress. The flow is allowed by dislocations and grain boundaries and it is a precursor of a microscopic plastic flow as in ordinary solids (e.g. ice) and as it is
probably observed in Ref. \cite{balibar}. A microscopic  yield-stress could be defined by the smallest gravity $\mathcal G$ such that no dislocations, defects nor grain boundaries appear. In the present model this is for  $\mathcal G <  10^{-4} $.

\noindent
In conclusion, we have shown a fully explicit model of supersolid that display either solid-like behavior or superflow depending on the external constrain and on the boundary conditions on the reservoir wall.
Our numerical simulations clearly show that, within the same model a nonclassical rotational inertia is observed as well a regular elastic response to external stress or forces without any flow of matter as in experiments \cite{chan,pressure}.


\begin{thebibliography}{99}
\bibitem{chan} E. Kim and M.H.W. Chan, a) Nature (London) {\bf 427}, 225 (2004); b)
Science {\bf 305}, 1941 (2004); c) Phys. Rev. Lett. {\bf 97}, 115302 (2006).
\bibitem{pressure} D. S. Greywall, Phys. Rev. {\bf B 16}, 1291 (1977); G. Bonfait, H. Godfrin, B. Castaing, J. Phys. (Paris) {\bf 50}, 1997 (1989); J. Day and J. Beamish, Phys. Rev. Lett. {\bf 96}, 105304 (2006).
\bibitem{pomric} Y. Pomeau and S. Rica, Phys.
Rev. Lett. {\bf 72}, 2426 (1994). 
\bibitem{pitgross} L.P. Pitaevski\v{\i}, Sov. Phys. JETP {\bf 13}, 451 (1961);
E.P. Gross, J. Math. Phys. {\bf 4}, 195 (1963).
\bibitem{bog} N.N. Bogoliubov, J. Phys. USSR {\bf 11}, 23 (1947). 
\bibitem{enz} T. Schneider and C.P. Enz, Phys. Rev. Lett. {\bf 27}, 1186  (1971).
\bibitem{son} D. T. Son,
Phys. Rev. Lett. {\bf 94}, 175301 (2005).
\bibitem{homogenization} A. Bensoussan, J.L. Lions, and G. Papanicolau, {\it asymptotic analysis in periodic Structures}, North-Holland Amsterdam (1978); E. S\'anchez-Palencia, {\it Non-Homogeneous Media and Vibration Theory}, Lecture Notes in Phys. {\bf 127} springer-Verlag, Berlin.
\bibitem{andreev} A.F. Andreev and I.M. Lifshitz, Sov. Phys. JETP.  {\bf 29}, 1107 (1969).
\bibitem{landau} L.D. Landau and E.M. Lifshitz, {\it Theory of Elasticity}, Pergamon Press (1980).
\bibitem{leggett} A.J. Leggett, Phys. Rev. Letters, {\bf
25}, 1543 (1970). 
\bibitem{milne} This problem describes an inviscid and incompresibble perfect fluid inside a rotating container see L.M. Milne-Thomson, {\it Theoretical Hydrodynamics}, Dover (1996), for exact solutions of various geometries in 2D see pages 263 to 271. In the context of superfluid this problem has been considered by A. Fetter, J. Low Temp. Phys. {\bf 16}, 533 (1974).
\bibitem{andreev2} A.F. Andreev, K. Keshishev, L. Mezov-Deglin, and A. Shalinikov, Zh. Eksp. Teor. Fiz. Pis'ma Red. {\bf 9}, 507 (1969) [JETP Lett. {\bf 9}, 306 (1969)].
\bibitem{balibar} S. Sasaki, R. Ishiguro, F. Caupin, H. J. Maris, S. Balibar, Science {\bf 313}, 1098 (2006).
\end{thebibliography}
\end{document}